\documentclass[final,5p,times,twocolumn,authoryear]{revtex4-1}
\usepackage{amssymb}
\usepackage{amsthm}
\usepackage{amsmath}
\usepackage{xcolor}

\usepackage{graphics}
\usepackage{epsfig}
\usepackage{subfigure}
\usepackage{rotating}

\usepackage{amsfonts}
\usepackage{amsmath}
\usepackage{amssymb}
\usepackage{placeins}
\usepackage{xcolor}
\usepackage{soul}
\usepackage{fixmath}
\usepackage[colorlinks]{hyperref}

\begin{document}

\title{Multipartite Entanglement in Bright Frequency Combs from Microresonators}

\author{Adrien Bensemhoun$^{1}$, C. Gonzalez-Arciniegas$^{2,3}$, Olivier Pfister$^{2}$, Laurent Labonté$^{1}$, Jean Etesse$^{1}$, Anthony Martin$^{1}$, Sébastien Tanzilli$^{1}$, Giuseppe Patera$^{3}$, Virginia D'Auria$^{1,*}$}

\affiliation{$^{1}$Université Côte d'Azur, CNRS, Institut de Physique de Nice (INPHYNI), UMR 7010, Parc Valrose, Nice Cedex 2, France}
\affiliation{$^{2}$University of Virginia Physics Department, 382 McCormick Rd, Charlottesville, VA 22903, USA}
\affiliation{$^{3}$Université Lille, CNRS, UMR 8523 - PhLAM - Physique des Lasers Atomes et Molécules, F-59000 Lille, France}

\email{Virginia.Dauria@univ-cotedazur.fr}

\begin{abstract}
We present a theoretical model of multimode quantum correlations in bright frequency combs generated in continuous-wave regime by microresonators above threshold. Our analysis shows how these correlations emerge from cascading four-wave mixing processes fed by the input pump as well as the generated bright beams. Logarithmic negativity criterion is employed to quantify entanglement between partitions of modes, demonstrating the transition from a bipartite regime just above the oscillation threshold to the multipartite one at higher input pump powers. Due to its generality, our model can be safely used to describe other kinds of non-linear $\chi^{(3)}$ cavities.
\end{abstract}


\maketitle

\section*{Introduction}
\label{introduction}

Silicon-based integrated photonics plays a central role in quantum optical technologies as it offers the possibility of generating, manipulating and detecting quantum states of light in high-density optical circuits~\cite{lipson2005guiding,banic2023integrated}. In this context, nonlinear microresonators on Silicon Nitride (SiN) have gained success in continuous-variable quantum optics~\cite{dutt2015chip, zhang2021squeezed}, as source of entanglement among modes at different optical frequencies generated by four-wave mixing (FWM). The association of SiN platform and spectral entanglement stand as a promising candidate for quantum computing~\cite{pfister2019continuous} and quantum communication~\cite{cozzolino2019high}. 
Many theoretical~\cite{gouzien2023hidden} and experimental~\cite{zhang2021squeezed,yang2021squeezed, zhao2020near} works have focused on quantum light from devices working below their oscillation threshold, with the demonstration of two-color squeezing in multiple beam pairs~\cite{yang2021squeezed,jahanbozorgi2023generation}. In this work, we focus on multimode features from bright frequency combs generated from a microresonator operating above threshold. The dynamics of such a regime is particularly interesting and largely investigated in classical optics, showing the appearance of a primary and of a secondary frequency comb, eventually leading to soliton production~\cite{karpov2019dynamics, chembo2016quantum}. In the quantum regime, theory~\cite{chembo2016quantum} and experiments~\cite{dutt2015chip} have shown twin beam-like intensity correlation among two colors of the primary comb. A signature of multimode behaviour has also been demonstrated in soliton microcombs via measurements of the second order photon correlation~\cite{guidry2022quantum} as well as the theoretical analysis of quadrature squeezing~\cite{guidry2023multimode}.

The scope of this theoretical paper is to show that multimode correlations are already present below the soliton threshold, in the simple case of the primary comb emitted by a continuous-wave-pumped microresonator: multimode features progressively arise from the cascade of subsequent FWM processes, where the signals initially produced by the degenerate conversion of the input continuous-wave (CW) pump act as seeds and/or as additional pumps for other (stimulated) FWM conversions that further feed the comb components (see Fig.~\ref{fig:FWM}). Our model considers a general FWM Hamiltonian, making our analysis easily extendable to other systems. Here, it is applied to a primary comb generated by a monochromatic pump from a microresonator operating above its oscillation threshold, providing the detail of modes' interaction in terms of the system quantum Langevin equations. Following an approach compliant to experimental verification, we characterize the entanglement in terms of the logarithmic negativity of the partial transpose of the covariance matrix~\cite{Adesso_2007}. We show the progressive transition from one-to-one correlations, similar to the one observed for below threshold systems~\cite{yang2021squeezed}, to richer multimode structures appearing when the cascaded FWMs become non negligible. By doing so, our work provides an intuitive and simple way to understand how (and why) we can leverage the rich dynamics above threshold for the generation of multimode bright quantum states for quantum technologies.

\section{Theoretical model}
\subsection*{FWM Hamiltonian and coupling matrix calculation}

\begin{figure}[h!]
\centering
\includegraphics[scale=0.21]{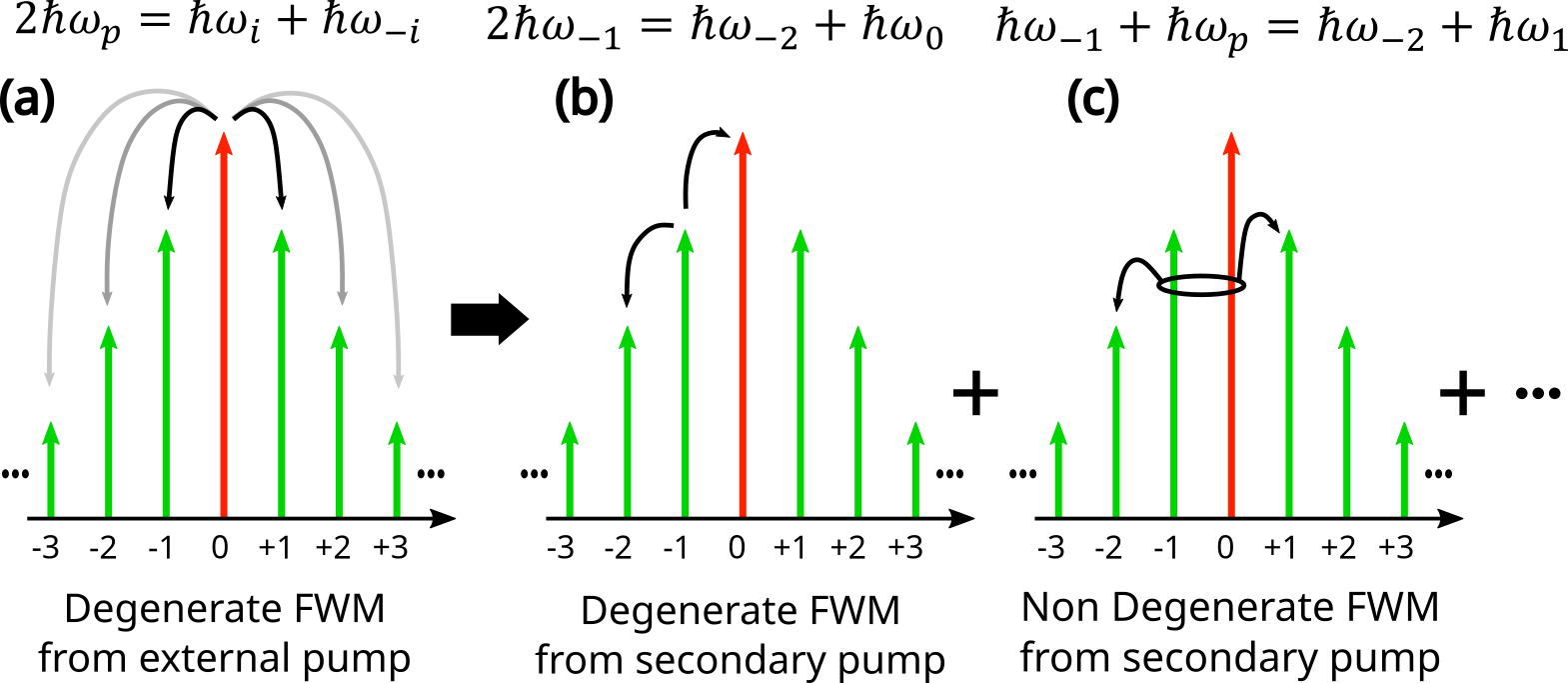}
\caption{Examples of FWM processes as primary combs. In red the laser resonant frequency mode. In green the frequency modes initially generated by degenerate FWM of the external pump (a) and that can act as pumps for other degenerate FWM (b) or non-degenerate FWM processes (c).}
\label{fig:FWM}
\hfill%    
\end{figure}

The starting point of the model is a very general FWM Hamiltonian describing the quantum dynamics of cavity-resonant frequency modes. Interacting modes are labelled as $n=0, \pm1, \pm2,...$, and their associated bosonic operators as $\hat{A}_n$ and $\hat{A}_n^{\dagger}$. As usual, they satisfy the boson commutation relations $[\hat{A}_n,\hat{A}_m^{\dagger}]=\delta_{n,m}$ and $[\hat{A}_n ,\hat{A}_m]=0$, $\delta_{n,m}$ being the Kronecker symbol~\cite{gardiner2004quantum, olivares2021}. The mode diagram is represented in Figure~\ref{fig:FWM}. Also note that the CW pump mode is referred to as $\hat A_0$. The considered interaction Hamiltonian is:
\begin{equation}
\label{hamilFWM}
    H=\displaystyle\sum_{klmn}^{} \: \delta_{k+l,m+n} \: \hat{A}^{\dagger}_{k} \: \hat{A}^{\dagger}_{l} \: \hat{A}_{m} \: \hat{A}_{n}.
\end{equation}
It describes in a general way the FWM process through which two photons in the frequency modes $k$ and $l$ are created from the annihilation of two photons in the frequency modes $m$ and $n$. The sum over all modes is justified by the fact that each frequency mode can combine and play the role of the pump for subsequent cascaded processes: no a priori choice is made on the pairs of photons that are annihilated, provided energy conservation ($\omega_k+\omega_l=\omega_n+\omega_m$), expressed by the Kronecker symbol, is respected. 

In the Heisenberg picture, the time evolution of modes' bosonic operators can be obtained as follows: 
\begin{equation} \label{shrodingereq}
\begin{split}
    &\frac{d\hat{A}_i}{dt}=\frac{i}{\hbar}[H,\hat{A}_i] \propto \hat{A}_k^{\dagger} \hat{A}_l^{\dagger} \hat{A}_m,\\
    &\frac{d\hat{A}_i^{\dagger}}{dt}=-\frac{i}{\hbar}[H,\hat{A}_i^{\dagger}] \propto \hat{A}_k^{\dagger} \hat{A}_l \hat{A}_m.
\end{split}
\end{equation}
The set of equations~\eqref{shrodingereq} exhibits terms in the form of a product of three annihilation and creation operators. Following a very standard procedure, such dynamic equations can be linearized by rewriting each bosonic operator as $\hat{A}_i=\alpha_i+ \hat{a}_i$ ($\hat{A}_i^{\dagger}=\alpha_i^\ast+\hat{a}_i^{\dagger}$), where $\alpha_i=\langle\hat A_i\rangle$, coinciding with the classical field amplitude, and $\hat{a}_i$ is the bosonic operator associated with the quantum fluctuations of the field in frequency mode $i$~\cite{gardiner2004quantum}. Note that by construction, $\langle \hat{a}_i \rangle=0$. Only terms at first order on $\hat{a}_i$ and $\hat{a}_i^{\dagger}$ are taken into account in the linearized equations. 
By doing so and moving in the interaction picture of the pump, the linearized system of analytical coupled equations can be conveniently written in a compact matrix form:
\begin{align}
\frac{d}{dt} 
\begin{pmatrix}
\boldsymbol{\hat{a}} \\
\boldsymbol{\hat{a}}^\dagger \\
\end{pmatrix}
= -4 i \cdot M_a
\begin{pmatrix}
\boldsymbol{\hat{a}} \\
\boldsymbol{\hat{a}}^\dagger \\
\end{pmatrix}
= -4 i
\begin{pmatrix}
F & G\\
-G^\ast & -F^\ast\\
\end{pmatrix}
\begin{pmatrix}
\boldsymbol{\hat{a}} \\
\boldsymbol{\hat{a}}^\dagger \\
\end{pmatrix},
\label{coupling}
\end{align}
where $\boldsymbol{\hat{a}}=\boldsymbol{\hat{a}}(t)$ stands for the column vector $\boldsymbol{\hat{a}}=(\hat{a}_{-K},...,\hat{a}_0,...,\hat{a}_{+K})^T$ with $K=\frac{N-1}{2}$ and N is an odd integer giving the number of modes. F and G are square matrices of dimension $2K+1$. The system of equations (\ref{coupling}) describes coupling between all the frequency modes, therefore $M_a$ must be Hamiltonian. This condition is verified using the relation: $(\Omega M_a)^T=\Omega M_a$ with $\Omega$ the symplectic matrix, $\Omega=
\begin{pmatrix}
\mathbb{O} & \mathbb{I}\\
-\mathbb{I} & \mathbb{O}\\
\end{pmatrix},
$
with $\mathbb{O}$ and $\mathbb{I}$ the zero and identity matrices. 
Matrix $F$ is Hermitian ($F=F^\dagger$) and includes FWM terms that are of the same kind as the parametric amplification. Matrix $G$ is symmetric ($G=G^T$) and takes into account self- and cross-phase modulation terms that are not included in $F$, as well as the modes' detuning from perfect cavity resonances that here will be taken as zero. The explicit expressions of $F$ and $G$ depend on the classical amplitudes
\begin{equation}
\begin{split} \label{fgmatrix}
&F_{kl}=\displaystyle\sum_{mn}^{} \: \delta_{k-l,m-n} \: \alpha_{m}\alpha_{n}^\ast, \\
&G_{kl}=\displaystyle\sum_{mn}^{} \: \delta_{-k-l,m+n} \: \alpha_{m}\alpha_{n}.
\end{split}
\end{equation}

\subsection*{Langevin equations and covariance matrix}

The dynamics of interaction modes inside the microresonator is described by the linearized Langevin equations, explicitly taking into account losses~\cite{gardiner2004quantum}. In the following we will write them in terms of the amplitude and phase quadratures $\hat{q}_i=\frac{1}{\sqrt{2}}(\hat{a}_i+\hat{a}_{i}^\dagger)$ and $\hat{p}_i=\frac{i}{\sqrt{2}}(\hat{a}_i-\hat{a}_{i}^\dagger)$, that are hermitian, measurable operators. The matrix 
\begin{equation}
V=\frac{1}{\sqrt{2}}%\cdot 
\begin{pmatrix}
\mathbb{I} & \mathbb{I}\\
-i%\cdot
\mathbb{I} & i%\cdot
\mathbb{I}\\
\end{pmatrix}
\label{vmatrix}
\end{equation}
allows performing the basis change from Eq.~\eqref{coupling} vector $(\boldsymbol{\hat{a}}(t),\boldsymbol{\hat{a}}^\dagger(t))^T$ to quadrature vector ${\boldsymbol{\hat{R}}}(t)=(\hat{q}_{-K}(t),..,\hat{q}_{+K}(t),\hat{p}_{-K}(t),..,\hat{p}_{+K}(t))^T$ . Accordingly, quadrature Langevin equations read as
\begin{align}
    \frac{d\boldsymbol{\hat{R}}}{dt} = (-\boldsymbol{\gamma}+M)\boldsymbol{\hat{R}}+\sqrt{2\boldsymbol{\gamma}}\boldsymbol{\hat{R}}_{\mathrm{in}}.
\label{langevin}
\end{align}
In this expression, $M=V^{-1}M_aV$ is the quadrature coupling matrix, $\boldsymbol{\hat{R}}_{\mathrm{in}}(t)$ is the quadrature vector of the resonator input modes and the matrix $\boldsymbol{\gamma}=\gamma \mathbb{I}$ represents the losses (assumed to be identical for all involved modes). Standard input-output relations~\cite{gardiner2004quantum} yield the quadratures of the fields at the cavity output 

\begin{align}
\boldsymbol{\hat{R}}_{\mathrm{out}}=\sqrt{2\gamma}\boldsymbol{\hat{R}}-\boldsymbol{\hat{R}}_{\mathrm{in}}.
\end{align}

\noindent Solutions of Eq.~\eqref{langevin} are found in the frequency domain by applying Fourier transform on the slowly varying envelopes:
\begin{align}
    \Tilde{\boldsymbol{R}}(\omega) = \frac{1}{\sqrt{2\pi}} \int_{-\infty}^{+\infty} e^{-i\omega t} \boldsymbol{\hat{R}}(t) dt.
\label{Rcomplex}
\end{align}
Note that quadrature operators $\Tilde{\boldsymbol{R}}(\omega)$ are conjugate symmetric with respect to the transformation $\omega\leftrightarrow - \omega$, $\hat{R}^ {\dag}(\omega)=\hat{R}(-\omega)$, so as to ensure operators' Hermiticity in time domain. In Eq.~\eqref{Rcomplex}, the analysis frequency $\omega \in \mathbb{R}$ labels the spectral components of modes' quantum noise, as retrieved, for instance, by a frequency homodyne. 
Note that, in the Fourier space, the quadratures of modes at the input and output of the resonator are connected via the transfer function matrix $\mathbf S(\gamma,\omega)$~\cite{gouzien2020morphing}

\begin{align}
\hat{\mathbf{R}}_{\mathrm{out}}(\omega)=\mathbf S(\gamma,\omega)\,\hat{\mathbf{R}}_{\mathrm{in}}(\omega).
\end{align}

\noindent The transfer function can be expressed as 
\begin{align}
   \mathbf S(\gamma,\omega)=\sqrt{2\boldsymbol{\gamma}}\,(i\omega\mathbb{I}+\boldsymbol{\gamma}-M)^{-1}\sqrt{2\boldsymbol{\gamma}}-\mathbb{I}.
\label{Somega}
\end{align}
\noindent To preserve the commutation rules, $\mathbf S(\gamma,\omega)$ satisfies the relation 
\begin{align}
\mathbf S(\gamma,\omega\,)\mathbf\Omega\, \mathbf S(\gamma,-\omega)^T=\mathbf\Omega
\end{align}

\noindent (see Ref.~\citenum{gouzien2020morphing} for details). Its explicit expression is a function of the cavity losses as well as of the stationary solutions of the system, $\{\alpha_i\}$, that can be obtained by solving the Lugiato-Lefever equations associated with the system above threshold~\cite{chembo2016quantum}.

\noindent The function $\mathbf S(\gamma,\omega)$ yields the noise covariance matrix of cavity output modes~\cite{Adesso_2007} 
\begin{align}
\mathbf \sigma(\gamma,\omega)=\frac{1}{2}\mathbf S(\gamma,\omega)\,\mathbf S^\dagger(\gamma,\omega).
\end{align}
The analytic form of $\sigma(\gamma,\omega)$ allows retrieving quadrature correlations between the different frequency modes. Remarkably, the covariance matrix is in general a smooth and complex function of $\omega$~\cite{gouzien2020morphing}. Note that the complex feature is usually shown in below-threshold silicon microresonators for $\omega \neq 0$~\cite{gouzien2023hidden}. In experiments, standard homodyne detection does not take into account asymmetry on spectral noise components between positive ($\omega$) and negative ($-\omega$) frequencies that can arise due to the imaginary part of the quadratures~\cite{barbosa2013beyond}. In other words, it only gives access to the real part of the quadrature, i.e. to the real part of the covariance matrix. Our analysis thus focuses on the real part of the covariance matrix only: this leads to a sub-optimal estimation of correlation~\cite{gouzien2023hidden} but allows keeping the analysis adherent to quantities that can actually be measured in the laboratory.  

\section{Logarithmic negativity criterion}

\begin{figure}[h!]
\centering
\includegraphics[scale=0.2]{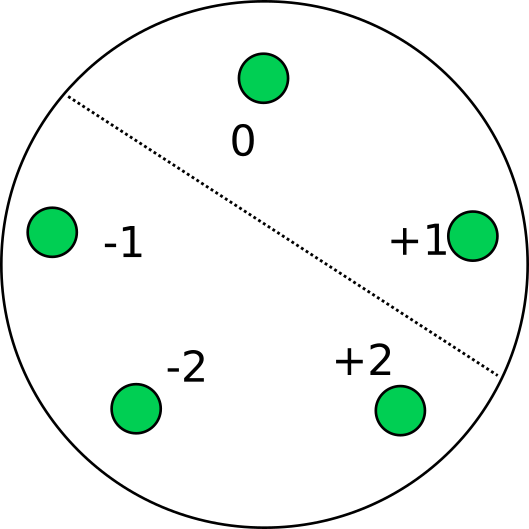}
\caption{Representation corresponding the bipartition: $\{-2,-1,2\}$:$\{0,1\}$ in the case of 5 modes. Such a partition corresponds to the partial transpose matrix $\Pi = \begin{pmatrix}
\mathbb{I} & \mathbb{O}\\
\mathbb{I} & \mathbb{D}\end{pmatrix}$ where $\mathbb{D}=diag(1,1,-1,-1,1)$.}
\label{fig:examplebipartition} 
\end{figure}

We investigate entanglement between the modes at the microresonator output in terms of the logarithmic negativity. Such a strategy relies on the analysis of the matrix $\sigma_{PT} = \Pi\sigma \Pi^{-1}$ corresponding to the partial transpose of the covariance matrix with respect to a bipartition of modes defined by partitioning operator $\Pi$. In the quadrature basis, this transformation can be implemented by simply inverting the sign of phase quadratures corresponding to the modes in one of the two partitions~\cite{simon2000peres}, see Fig.~\ref{fig:examplebipartition} as an example. 

In the continuous-variable regime, the logarithmic negativity is defined as
\begin{equation}
    \Sigma = - \displaystyle\sum_{i:n_i<1} \ln{n_i},
    \label{logneg}
\end{equation}
where $\{n_i\}$ is the set of symplectic eigenvalues of $\sigma_{PT}$ as obtained by diagonalizing the matrix $|i\Omega \sigma_{PT} |$ and normalizing them to those corresponding to vacuum state (i.e., to a diagonal covariance matrix whose elements are given by the shot-noise level)~\cite{Adesso_2007}. Note that, as justified above, in what follows we rather examine the eigenvalues of $|i\Omega \operatorname{Re}(\sigma_{PT}) |$. 

In Eq.~\eqref{logneg}, the condition $n_i<1$ implies that the summation is restricted to the symplectic eigenvalues $n_i$ that are smaller than 1, i.e., associated with the presence of entanglement~\cite{vidal2002computable}. According to the PPT criterion, a positive $\Sigma$ will thus indicate the presence of entanglement between the two parts of the chosen partition. The logarithmic negativity provides a necessary and sufficient condition when the mode partition is in the form 1x(N-1)~\cite{adesso2004quantification}: for these cases $\Sigma$ can be used to quantify genuine entanglement. Note that, as, in general, $\sigma(\gamma,\omega)$ at the microresonator output is not a bi-symmetric matrix, for a generic bipartition of the form $L\times(N-L)$, with $1\le L<N$, a positive $\Sigma$ only provides a sufficient condition for entanglement~\cite{adesso2004quantification} but can be used as an entanglement witness. 

\section{Results}

\subsection*{Multimode features in the primary comb}

\begin{figure}[h!]
\centering
\includegraphics[scale=0.30]{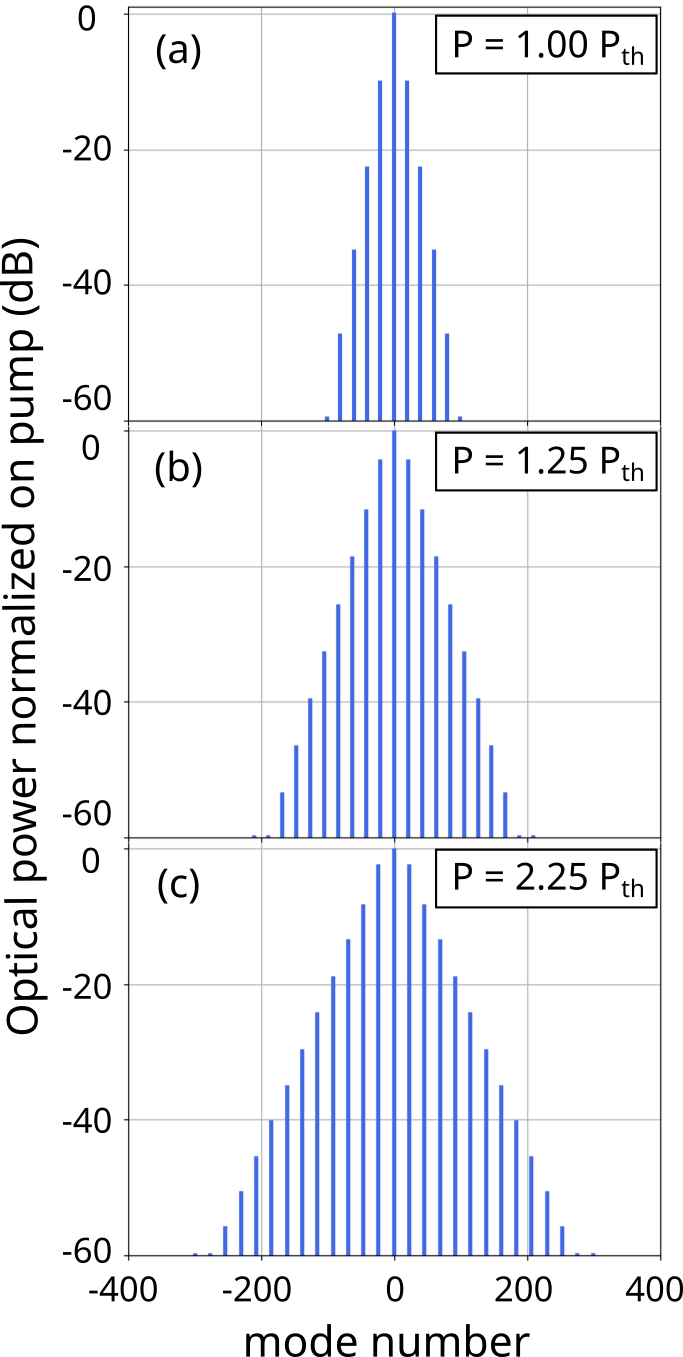}
\caption{Frequency primary combs for $\omega=0$ and for 3 laser pump powers: (a): P=1.00\,P$_{th}$, (b): P=1.25\,P$_{th}$, (c): P=2.25\,P$_{th}$. The power is expressed in dB as a function of the mode number: it corresponds to $|\alpha_i|^2$. The detuning from the cold cavity resonance wavelength is set to $0\,$nm.}
\label{fig:pumppower}   
\end{figure}

To highlight the effect of cascaded FWM processes on multimode correlations, we now consider different pump powers. As an example, Fig.~\ref{fig:pumppower} shows the classical relative intensity $|\alpha_i|^2/|\alpha_0|^2$ of primary comb components as obtained from numerical simulations of Lugiato-Lefever equations by injecting different input powers in mode $m=0$, i.e., P=1.00\,P$_{th}$, P=1.25\,P$_{th}$, and P=2.25\,P$_{th}$, where P$_{th}$ is the microresonator parametric oscillation threshold. The other parameters of the simulation are the pump detuning with respect to a given cavity cold resonance and second order dispersion. They are $\Delta_{\mathrm{p}}=0$ and $\Omega_2=-0.01\gamma$, respectively. As can be seen, the pump power P has a strong effect on the stationary solutions, $\{\alpha_i\}$, leading, as expected, to the progressive excitation of an increasing number of modes. Note that for zero pump detuning on as considered here, no secondary comb is observed; the theoretical investigation of the secondary comb regime has already be performed elsewhere~\cite{chembo2016quantum,guidry2023multimode} and it is beyond the scope of this work.

In what follows, we apply the logarithmic negativity criterion to the $\operatorname{Re}[\sigma(\gamma,\omega)]$ of a primary comb of 5 modes ($0,\pm1,\pm2$). This is the simplest multimode configuration after the case of 3 modes pump, $\pm1$, that has already been studied theoretically~\cite{chembo2016quantum} and experimentally~\cite{dutt2015chip} for twin beam-like correlations. Analysing the features of 5 modes is thus interesting to understand how quickly the system behaves as a multimode entanglement source. A discussion on the results obtained with a higher number of modes is provided in the last section of this work.

Multimode entanglement can be studied by plotting the logarithmic negativity as a function of the analysis frequency when normalizing for simplicity the cavity losses to 1 ($\gamma=1$). We start by looking for entanglement in the case of $\omega=0$. Correspondingly, the covariance matrix of modes' quadrature fluctuations is fully real and, as a consequence, entirely accessible to experiments by means of standard homodyne detections~\cite{gouzien2023hidden, barbosa2013beyond}. The logarithmic negativity has been evaluated for all the bipartitions with N=5. Table~\ref{fig:resultw0} summarizes the results for $\omega=0$ at different input pump powers, by taking into account the fact the role of interaction modes is symmetrical with respect to the CW pump mode (mode $0$).

\begin{table}[htbp]
\centering
\begin{tabular}{|p{3cm}|c|c|c|c|}
\hline
\multicolumn{1}{|c|}{\raisebox{-2.5ex}{\Large Partition}} & \multicolumn{3}{c|}{\raisebox{-0.9ex}{\Large $\Sigma$}} \\
\cline{2-4}
& \rule{0pt}{0.45cm}\raisebox{0.05cm} {1.00\,P$_{th}$} & \raisebox{0.05cm} {1.25\,P$_{th}$} & \raisebox{0.05cm} {2.25\,P$_{th}$} \\
\hline
\rule{0pt}{0.40cm} \centering $\{-2\}$:$\{-1,0,1,2\}$ & 1.21 & 1.17 & 0.99 \\
\centering $\{-1\}$:$\{-2,0,1,2\}$ & 1.22 & 1.32 & 1.23 \\
\centering $\{0\}$:$\{-2,-1,1,2\}$ & 0.15 & 0.91 & 1.29 \\
\centering $\{-2,-1\}$:$\{0,1,2\}$ & 1.26 & 1.48 & 1.37 \\
\centering $\{-2,0\}$:$\{-1,1,2\}$ & 1.21 & 1.18 & 1.21 \\
\centering $\{-2,1\}$:$\{-1,0,2\}$ & 1.21 & 1.21 & 1.14 \\
\centering $\{-2,2\}$:$\{-1,0,1\}$ & 0.11 & 0.64 & 0.90 \\
\centering $\{-1,0\}$:$\{-2,1,2\}$ & 1.22 & 1.33 & 1.32 \\
\rule{0pt}{0.30cm} \centering \raisebox{0.2ex}{$\{-1,1\}$:$\{-2,0,2\}$} & \raisebox{0.2ex}{0.18} & \raisebox{0.2ex}{0.99} & \raisebox{0.2ex}{1.17} \\
\hline
\end{tabular}
\caption{Summary of logarithmic negativity for every bipartitions for $\omega=0$. It is shown that the state cannot be separated in any way.}
\label{fig:resultw0}
\end{table}

%\begin{figure}[h!]
%\centering
%\includegraphics[scale=0.30]{Images/sum_multimode.png}
%\caption{Summary of logarithmic negativity for every bipartitions for $\omega=0$. It is shown that the state cannot be separated in any way.}
%\label{fig:resultw0}
%\hfill%    
%\end{figure}

As shown in the table, $\Sigma$ is always $>0$, indicating the presence of entanglement whatever bipartition and pump powers are considered, thus providing a simple intuition of multimode correlations. Entanglement is found, in particular, in the case of all 1x(N-1) bipartitions, for which, as discussed, the value of $\Sigma$ also stands as an entanglement quantifier. In the following discussions for $\omega\neq0$, we will only focus on partitions of this kind. 

\subsection*{From two-mode to multimode correlations}

With the appearance of cascaded FWM processes made possible by higher $\{\alpha_i\}$, entanglement becomes progressively more and more multimode. A signature of this can be seen by computing $\Sigma$ for bipartitions of the kind 1xL, with L$\leq5$. This formally corresponds to start from the 5 mode-model and to subsequently trace out a certain number of modes from the calculation of $\sigma(\gamma, \omega)$, i.e. to consider only some chosen modes of the comb, while disregarding the others. For the cases with a number of modes smaller than 5, the most entangled bipartition is taken.

\begin{figure}[h!]
\centering
\includegraphics[scale=0.4]{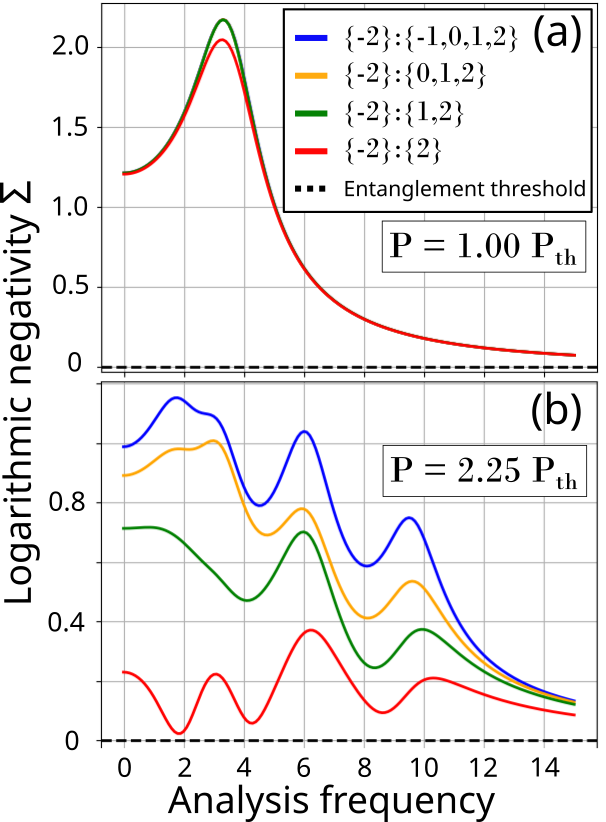}
\caption{Logarithmic negativity as a function of the analysis frequency $\omega$ for the partitions -2xM with 5 modes $\{-2\}$:$\{-1,0,1,2\}$ (blue), 4 modes $\{-2\}$:$\{0,1,2\}$ (orange), 3 modes $\{-2\}$:$\{1,2\}$ (green) and 2 modes $\{-2\}$:$\{2\}$ (red). The power is: (a): P=1.00\,P$_{th}$, (b): P=1.25\,P$_{th}$ and (c): P=2.25\,P$_{th}$.}
\label{fig:Trace_-2}
\hfill%    
\end{figure}

Following an approach similar to what has been done in above threshold experiments~\cite{dutt2015chip}, we start by analysing the simplest case of correlations between paired modes $\{-i\}$:$\{i\}$: these are the modes originally generated by the primary process of degenerate FWM of the input pump. Simulation results are shown in Fig.\ref{fig:Trace_-2} for partitions of mode $-2$ with the others (analogous results are obtained when considering for instance partitions of mode $-1$). 

\begin{figure}[h!]
\centering
\includegraphics[scale=0.11]{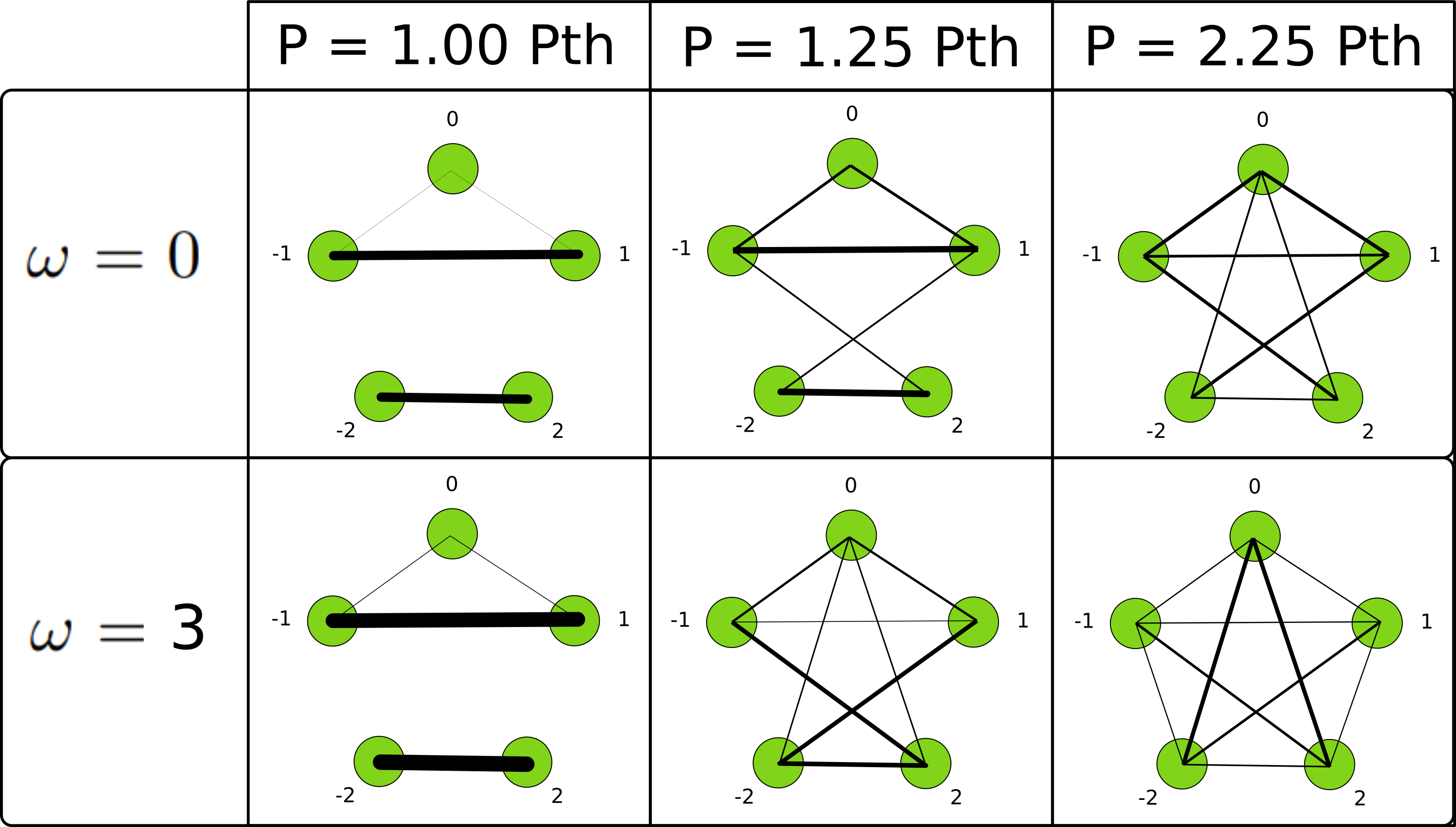}
\caption{Logarithmic negativity values symbolized by link thickness between the partitions of the two considered modes for $\omega=0$, $\omega=3$ and for the power P=1.00\,P$_{th}$, P=1.25\,P$_{th}$ and P=2.25\,P$_{th}$.}
\label{fig:graph2modes}
\hfill%    
\end{figure}

As a first general remark, note that, as expected~\cite{gouzien2020morphing}, the logarithmic negativity between $\{-i\}$:$\{i\}$ ($\{-2\}$:$\{2\}$ in the figure) depends on the analysis frequency in a smooth (although non trivial) way, tending to zero when the analysis frequency goes well beyond the cavity bandwidth. Moreover, the quantum noise of emitted states reaches coherent vacuum's levels. Interestingly, in multiple configurations, highest $\Sigma$ are found for $\omega\neq 0$, despite the covariance matrix is not real.
Figure~\ref{fig:Trace_-2}-(a) also shows that at low pump powers, entanglement between paired symmetrical modes $\{-2\}$:$\{2\}$ corresponds to an optimal $\Sigma$ (and analogously for $\{-1\}$:$\{1\}$, not in the figure). Remarkably, the logarithmic negativity does not change significantly with the progressive increase of the number of comb modes (i.e., when comparing results obtained for the partitions $\{-2\}$:$\{2\}$, $\{-2\}$:$\{0,1,2\}$ or $\{-2\}$:$\{-1,0,1,2\}$). This indicates that entanglement involving mode $-2$ is mainly due to its bipartite correlation with its symmetric twin $+2$. The situation changes for the power P=2.25\,P$_{th}$, when cascaded FWM processes become non negligible. The entanglement of the partition $\{-2\}$:$\{2\}$ decreases with P, due to the fact that when the amplitude and the number of modes in the bipartition increases, bipartite entanglements of paired symmetrical modes $\{-i\}$:$\{ i\}$ deteriorate to the benefit of multimode quantum correlations. Correspondingly, optimal $\Sigma$ increases when including in the model a higher number of considered mode, thus suggesting that each of the 5 mode somehow shares entanglement with mode $-2$. This is confirmed by the progressive appearance of bipartite entanglement in non-symmetric partitions $\{i\}$:$\{j\}$ (as for example, $\{-2\}$:$\{1\}$, $\{-2\}$:$\{0\}$ and $\{-2\}$:$\{-1\}$) for P$>$1.00\,P$_{th}$). Such a transition from a bipartite to a multipartite regime is shown in Fig.~\ref{fig:graph2modes} for every 2-mode bipartitions. Higher $\Sigma$ values are represented by a thicker link between the two considered modes. As expected, for P=1.00\,P$_{th}$ entanglement is mostly localized around twin modes $\{-1\}$:$\{1\}$ and $\{-2\}$:$\{2\}$ corresponding to bipartite entanglement as it can be found below the oscillation threshold ~\cite{yang2021squeezed,jahanbozorgi2023generation}. When increasing the pump power, other links are created between non-symmetrical modes. It reveals that other modes like $+1$ or $-1$ play the role of pumps and contribute to multimode entanglement formation. For P=2.25\,P$_{th}$ links between non-symmetrical modes becomes stronger than those between symmetrical modes and the system makes a transition from bipartite to multipartite entanglement.

\subsection*{Spectral comparison between 3-, 5-, and 7-mode models}

\begin{figure}[t]
\centering
\includegraphics[scale=0.4]{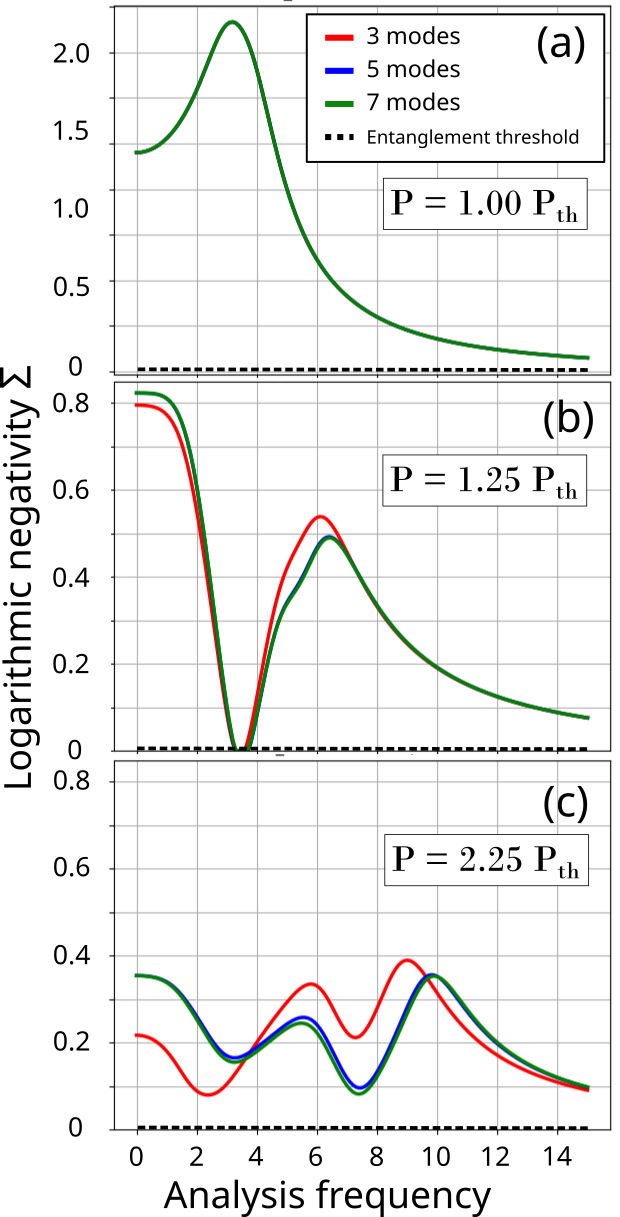}
\caption{Logarithmic negativity of the bipartition $\{-1\}$:$\{1\}$ as a function of the analysis frequency $\omega$ for P=1.00\,P$_{th}$ (a), P=1.25\,P$_{th}$ (b), P=2.25\,P$_{th}$ (c). The model takes into account the interaction between 3 modes (red), 5 modes (blue) and 7 modes (green).} 
\label{fig:2modes}
\hfill%    
\end{figure}

Results presented so far consider a primary frequency comb of only 5 modes. However, as seen from Fig.~\ref{fig:pumppower}, increasing the pump power leads to the excitation of a higher number of modes including at the powers chosen here to run the simulations. It is thus pertinent to verify the influence of neglected modes ($\pm 3, \pm 4,...$) in the entanglement analysis. To visualise the influence of the number of interacting modes we compute the logarithmic negativity of the bipartition $\{-1\}$:$\{1\}$ starting from a model with 3, 5 and 7 modes in Hamiltonian~\eqref{hamilFWM}. Results are plotted in figure~\ref{fig:2modes}. For a pump power P=1.00\,P$_{th}$, including further modes in the model does not significantly affect the logarithmic negativity. This is in agreement with what expected from the analysis of bipartite entanglement previously discussed, confirming that nearby the threshold most of entanglement is due to one-to-one correlational among symmetrical modes. As reasonable, differences arise when considering higher systems excitation levels, leading to differences between a three mode model (clearly insufficient) and the other one with higher modes. Nevertheless, only minor differences arise between the models with 5 and 7 modes for the pump values considered in this work. 

\section*{Summary and conclusions}

In this theoretical study we analyse multimode quantum correlations in bright frequency combs generated by microresonators operating above threshold. Our simple model is sufficient to understand how these correlations arise in the presence of cascading FWM processes. We use the logarithmic negativity criterion to quantify entanglement between partitions of modes and examined the influence of pump power and analysis frequency on entanglement. Our results show that at low pump powers, entanglement is mostly present between symmetrically paired modes as it can be found below the oscillation
threshold. As the pump power increases, cascaded FWM processes become more important, leading to the emergence of multipartite entanglement.
Our study sheds light on the complex dynamics of microresonators operating above threshold, highlighting their potential for generating high-dimensional multimode quantum states. Understanding and controlling these quantum correlations are essential steps towards harnessing the full potential of microresonator-based quantum technologies. We believe that our model can invite experimental realizations and future applications involving bright frequency comb from silicon based structures as key technological resources.

\section*{Acknowledgements}

This work has been conducted within the framework of the project SPHIFA (ANR-20-CE47-0012). V.D’A. acknowledges financial support from the Institut Universitaire de France (IUF)

\bibliography{bibliography}

\end{document}